# Analysis of Regional Cluster Structure By Principal Components Modelling In Russian Federation

**Aleksandr V. Bezrukov**

Statistics department

Plekhanov Russian University of Economics

Moscow, Stremyanny pereulok 36

Bezrukov.AV@rea.ru

## Abstract

In this paper it is demonstrated that the application of principal components analysis for regional cluster modelling and analysis is essential in the situations where there is significant multicollinearity among several parameters, especially when the dimensionality of regional data is measured in tens.

The proposed principal components model allows for same-quality representation of the clustering of regions. In fact, the clusters become more distinctive and the apparent outliers become either more pronounced with the component model clustering or are alleviated with the respective hierarchical cluster. Thus, a five-component model was obtained and validated upon 85 regions of Russian Federation and 19 socio-economic parameters. The principal components allowed to describe approximately 75% of the initial parameters variation and enable further simulations upon the studied variables.

The cluster analysis upon the principal components modelling enabled better exposure of regional structure and disparity in economic development in Russian Federation, consisting of four main clusters: the few-numbered highest development regions, the clusters with mid-to-high and low economic development, and the 'poorest' regions. It is observable that the development in most regions relies upon resource economy, and the industrial potential as well as inter-regional infrastructural potential are not realized to their fullest, while only the wealthiest regions show highly developed economy, while the industry in other regions shows signs of stagnation which is scaled further due to the conditions entailed by economic sanctions and the recent Covid-19 pandemic. Most Russian regions are in need of additional public support and industrial development, as their capital assets potential is hampered and, while having sufficient labor resources, their donorship will increase.

## 1 Introduction

The social differentiation and disparity, as well as interregional disparity, is one of the central problems of the developed modern society. Of the development goals set by the United Nations Organization that must be achieved until 2030, we should point out the "Reduction of inequality within and between countries" [1].

It is hardly arguable that socio-economic disparity is one of the sources of instability of the territorial development of a country, and is one of the struggles to achieve sustainable development of its economy. The President of Russia, V.V. Putin, has stated in his message to the Federal Assembly in 2018: "a person, his present and future, is the main meaning and goal of our development", which then further resulted in the May Decree to preserve and increase the human capital. The Decree specifies a goal of reducing the poverty level in the Russian Federation by half, which, in the opinion of V.V. Putin, could enable a breakthrough in the development of the country and improve the life quality of its citizens.

It therefore becomes necessary to perform in-depth exploration of regional data to expose the possible problems in socio-economic development and disparity in regional development, and to improve the decisions on both local and federal level.

The regional modelling and forecasting is, evidently, one of the primary foci of efficient modern governing and decision-making for public organizations and the government in overall. It is essential for the government to provide integral infrastructure and innovative development with collaborating regions, putting geographic adjacency to good use in economic and financial development, being able to adequately assess and forecast the economic situation among regions. Considering the Russian Federation territorial diversity and spatiality, it would be necessary to apply a whole variety of economic characteristics in a model while attempting to simulate the regional development as a system, to highlight and play to its strengths and determine the points of growth.

In author's opinion, the relevance of dimensionality reduction modelling methods in this situation becomes most apparent, as a large variety of economic characteristics of regional development and the number and diversity of Russian regions would produce mixed results in clustering and cluster modelling methods of analysis. It appears that the principal component modelling would be a very appropriate method to simulate the economic regional development, as the principal components would allow to achieve the following theoretical and practical purposes:

- reveal the cluster structure of regions in the dissection of economic development and the disparity existing among the regions;

- highlight the main factors influencing the interregional variation of sustainable growth characteristics;

- allow to obtain a model of behavior of the regional characteristics of their socio-economic development, to forecast their development;

- gain insight into the disparity causes among the regions in terms of sustainable economic growth and draw the possible suggestions to certain aspects that require tackling.

## 2  Simulations

The presented paper focuses on obtaining a component model of Russian regional development upon the system of 19 economic indicators of activity by 85 regions:

Table 1: System of indicators for the clustering of regional development of Russian Federation

| | **Indicator** |
|---|---|
| 1. | GRP per capita, rubles |
| 2. | The volume of investments in fixed assets, million rubles |
| 3. | Cost of fixed assets, million rubles |
| 4. | Expenditures on technological innovations, million rubles |
| 5. | Industrial producer price index, |
| 6. | Average per capita cash income, rubles |
| 7. | The share of the population with cash incomes below the subsistence level, in |
| 8. | Gini coefficient, at times |
| 9. | Consumer Price Index |
| 10. | Cost of a fixed set of consumer goods and services, rub |

11. Coefficients of migration growth per 10 000 population
12. Population change
13. Demographic load factors, per 1000 people of working age
14. Natural population growth rates per 1000 people
15. Life expectancy at birth, years
16. Number of labor resources, thousand people
17. The share of persons under working age employed in the economy in the total number of employed
18. Unemployment rate
19. Real accounted wages of employees of organizations

The initial data were aggregated from official sources by Rosstat, while the missing values for certain regions replaced by variable mean levels, and the dataset was subsequently normalized for purposes of principal component modelling by means of the R package.

In its own term, the principal component modelling can be considered in the following stages.

1) Understanding the structure of the data and the variables;

2) Obtaining the component model and determining its parameters;

3) Evaluating the quality of the model and interpreting it.

The parallel coordinates plot (Fig. 1) and the heatmap (Fig.2) allow to state the following findings.

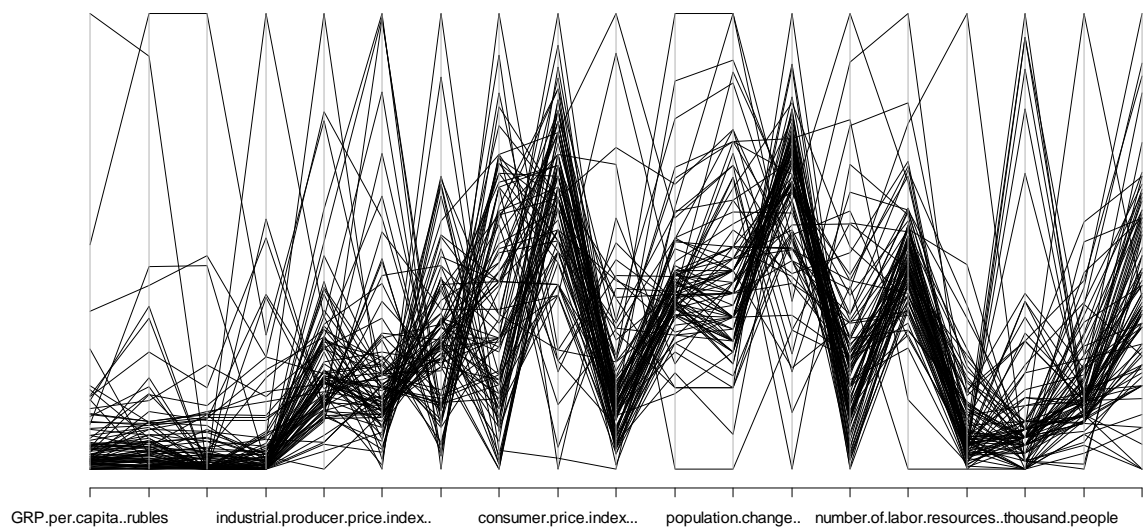

Figure 1: Parallel coordinates graph of the regional development parameters

- the structure of variables has clear clusterization, while the clusters of regions themselves have somewhat mixed and unclear composition in their correspondence to the variables behavior; thus, heuristic methods of cluster searching by density would not produce meaningful results;

- there exists strong multicollinearity among arrays of variables;

- a more detailed analysis of variables clustering by applying the complete linkage algorithm determines and confirms four distinctive groups of parameters (Fig.3).

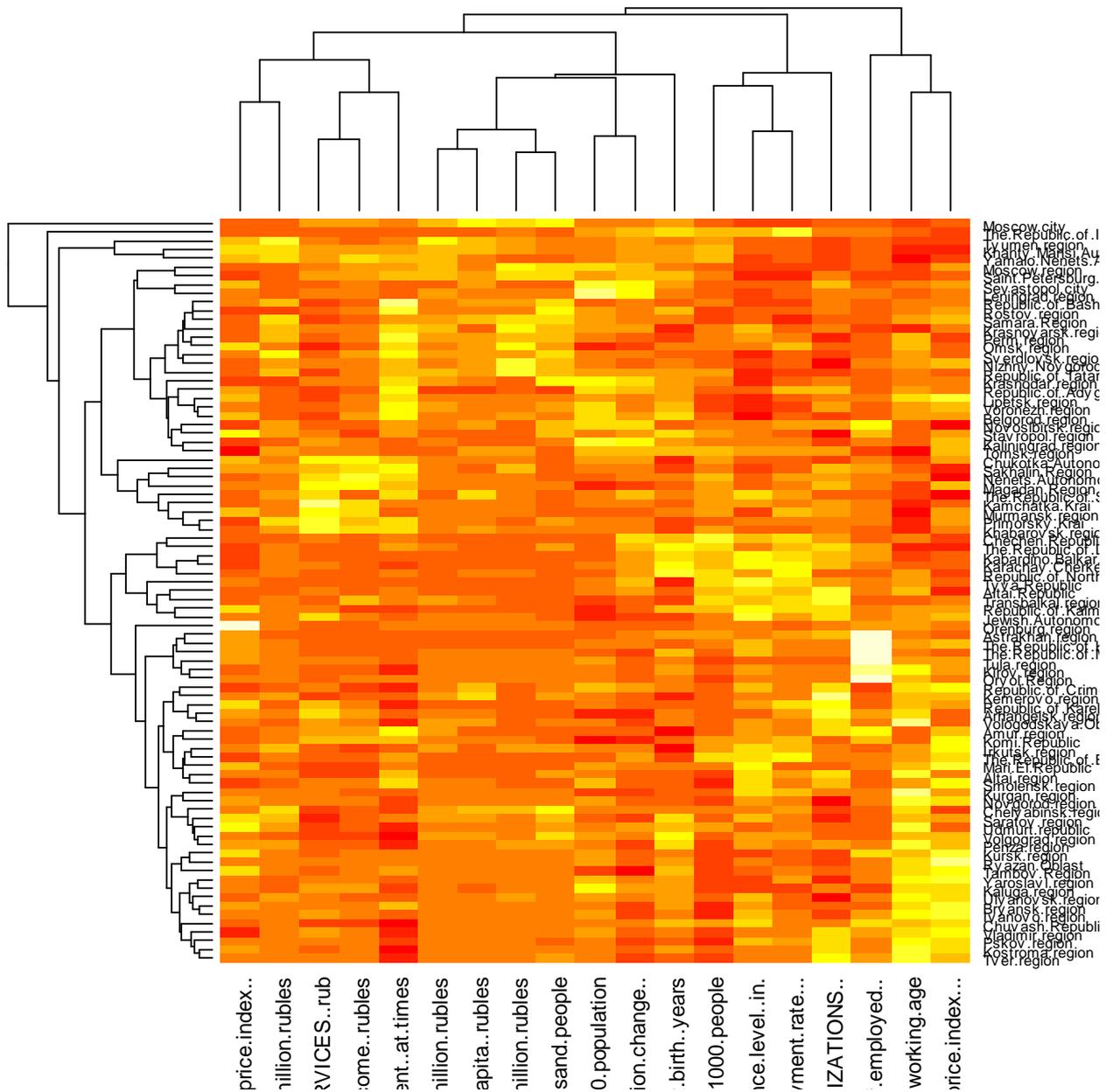

Figure 2: Heatmap plot of regional development parameters

The subsequent stage would be to perform the principal components modelling. Let us remind that the essence of the principal components is to reduce the dimensionality of the initial dataset to a new set of variables, each being a linear function of the initial parameters, with the following properties, essential to the goal of the present task:

- the principal components are uncorrelated and represent a coordinate system of orthogonal axes;

- the principal components describe the behavior of the initial variables, rather than individual observations;

- the principal components factor loadings are the correlations with the initial variables which are considered as the effect of these higher-order factors; thus, principal components are interpreted upon the factor loadings.

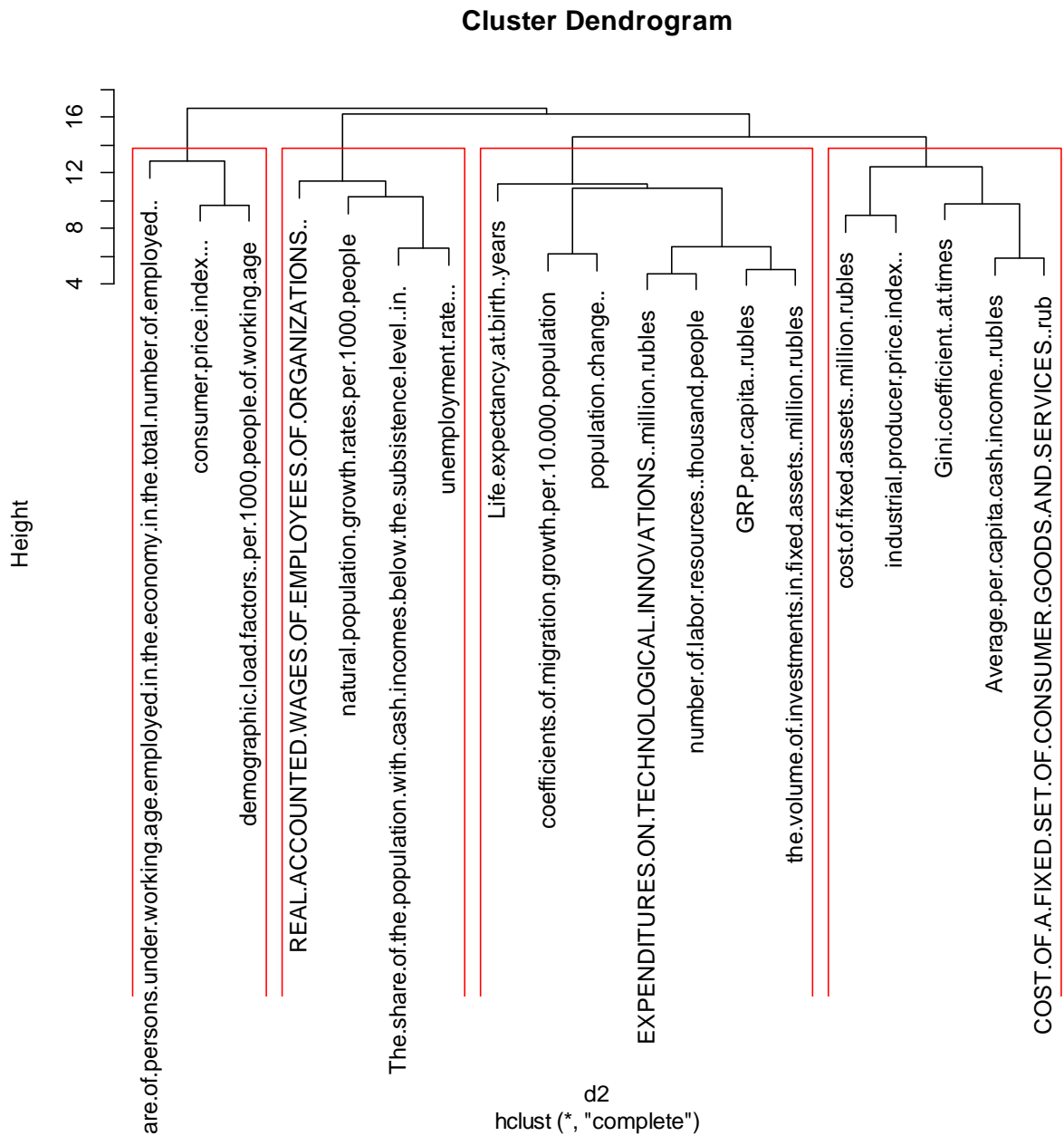

Figure 3: Cluster structure of regional development variables

Having performed the principal component analysis, it was necessary to decide upon the number of principal components to be considered, which has been done upon the scree criterion (Fig. 4) and the eigenvalues / explained variance percentage (Table 2). For principal components modelling the general guideline is to select the ones with eigenvalues higher than 1, while at the same time considering the scree plot inclination. Thus, five principal components were selected, describing 75,18% of behavior of the initial variables (see Table 2).

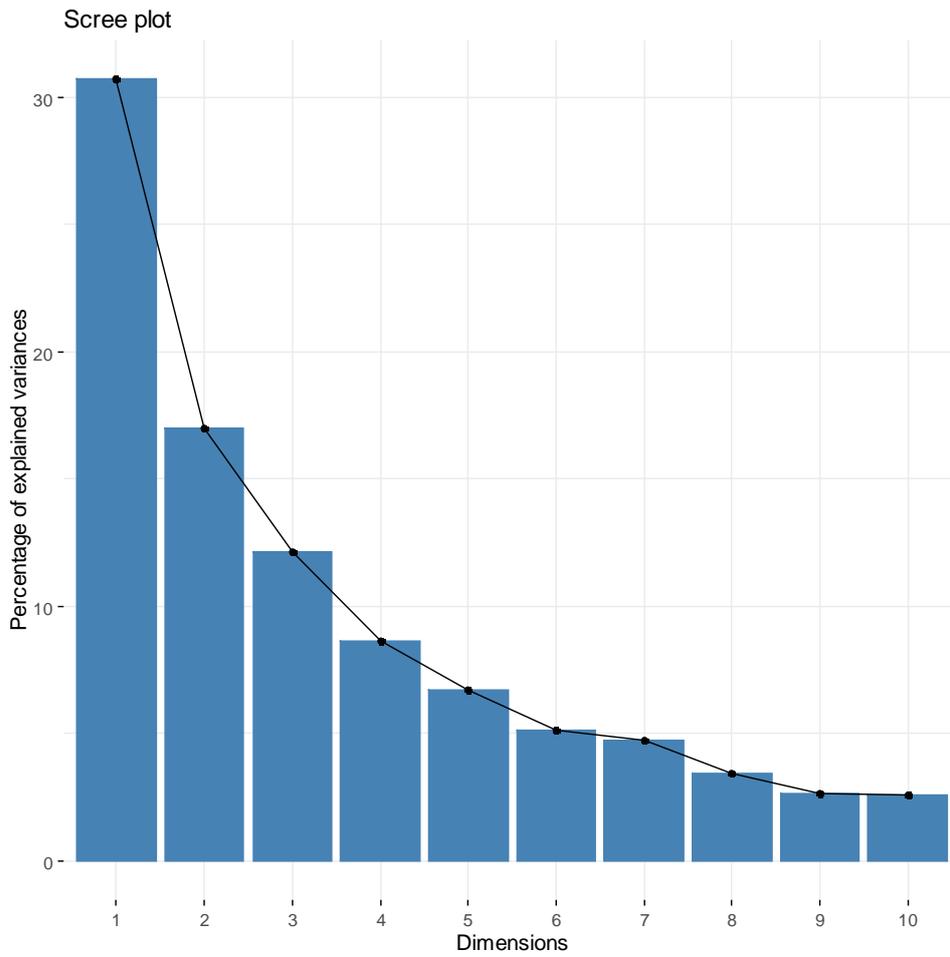

Figure 4: Scree plot for principal components model

Table 2: Eigenvalues of principal components

|        | eigenvalue  | variance percent | cumulative percent |
|--------|-------------|------------------|--------------------|
| Dim.1  | 5.832579887 | 30.69778888      | 30.69779           |
| Dim.2  | 3.232571219 | 17.01353273      | 47.71132           |
| Dim.3  | 2.300940659 | 12.11021399      | 59.82154           |
| Dim.4  | 1.644107304 | 8.65319634       | 68.47473           |
| Dim.5  | 1.273934146 | 6.70491656       | 75.17965           |
| Dim.6  | 0.970767333 | 5.10930176       | 80.28895           |
| Dim.7  | 0.896241187 | 4.71705888       | 85.00601           |
| Dim.8  | 0.649316999 | 3.41745789       | 88.42347           |
| Dim.9  | 0.502682237 | 2.64569599       | 91.06916           |
| Dim.10 | 0.487771042 | 2.56721601       | 93.63638           |
| Dim.11 | 0.365864420 | 1.92560221       | 95.56198           |
| Dim.12 | 0.234788679 | 1.23572989       | 96.79771           |
| Dim.13 | 0.179103669 | 0.94265089       | 97.74036           |
| Dim.14 | 0.129516751 | 0.68166711       | 98.42203           |
| Dim.15 | 0.103789464 | 0.54626034       | 98.96829           |
| Dim.16 | 0.083736033 | 0.44071596       | 99.40901           |
| Dim.17 | 0.076707881 | 0.40372569       | 99.81273           |
| Dim.18 | 0.030332739 | 0.15964599       | 99.97238           |
| Dim.19 | 0.005248349 | 0.02762289       | 100.00000          |

The factor loadings in the space of the first and the second principal axes are presented on Fig. 5., and on the biplot with factor scores on Fig. 6.

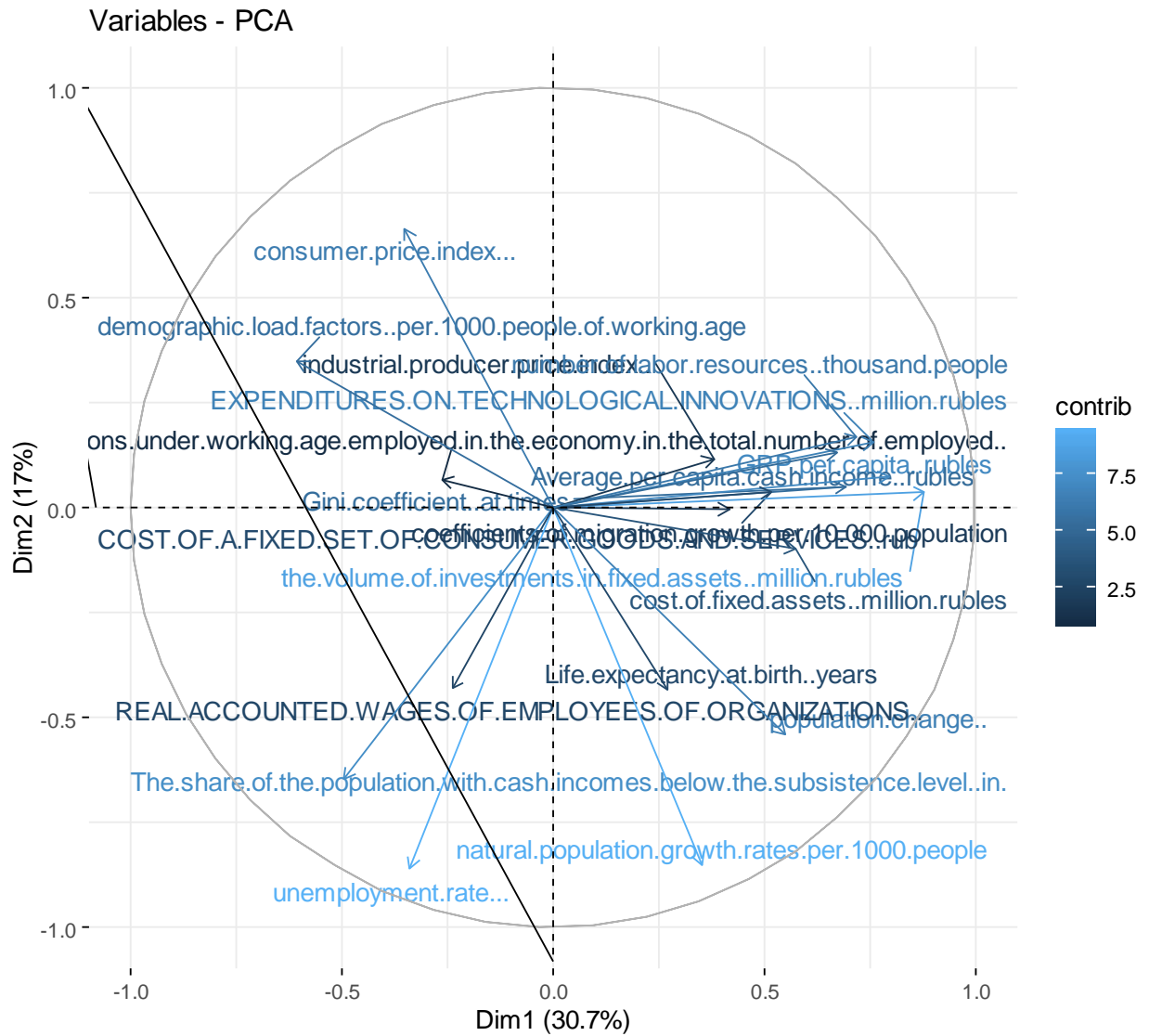

Figure 5: Factor loadings plot for the principal components model

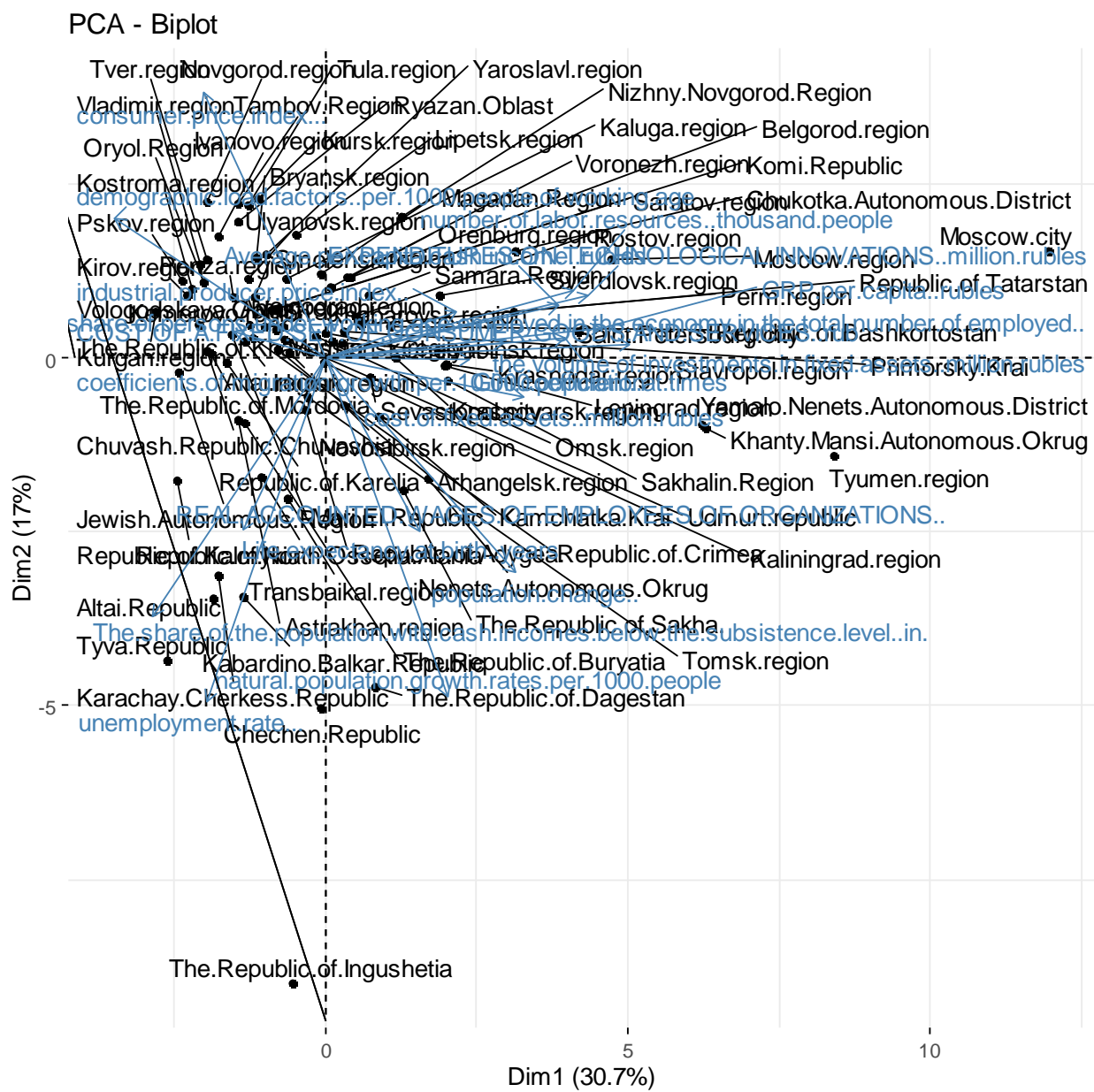

Figure 6: Biplot for factor scores and factor loadings for Russian Federation regions

## 3 Application

The coefficients for the principal axes are presented in the Table 3, representing the factor loadings of the component model.

Table 3: Factor loadings (unrotated) for regional development parameters

| Indicator Component | f1 | f2 | f3 | f4 | f5 |
|---|---|---|---|---|---|
| GRP per capita, rubles | 0.336 | 0.163 | 0.225 | 0.095 | 0.085 |
| The volume of investments in fixed assets, million rubles | 0.365 | 0.182 | 0.166 | 0.022 | -0.098 |
| Cost of fixed assets, million rubles | 0.239 | 0.161 | 0.04 | -0.315 | -0.381 |
| Expenditures on technological innovations, million rubles | 0.324 | 0.084 | 0.265 | 0.131 | 0.195 |
| Industrial producer price index, | 0.167 | 0.004 | 0.015 | -0.1 | -0.514 |
| Average per capita cash income, rubles | 0.303 | -0.118 | -0.374 | 0.055 | 0.029 |
| The share of the population with cash incomes below the subsistence level, in | -0.253 | 0.332 | -0.04 | -0.138 | 0.093 |
| Gini coefficient, at times | 0.302 | -0.057 | -0.162 | -0.146 | 0.107 |
| Consumer Price Index | -0.108 | -0.425 | 0.219 | 0.292 | 0.016 |
| Cost of a fixed set of consumer goods and services, rub | 0.235 | -0.101 | -0.444 | 0.063 | 0.108 |
| Coefficients of migration growth per 10 000 population | 0.032 | 0.1 | -0.25 | 0.562 | -0.063 |
| Population change | -0.057 | 0.193 | -0.093 | 0.577 | -0.333 |
| Demographic load factors, per 1000 people of working age | -0.237 | -0.16 | 0.399 | 0.114 | 0.018 |
| Natural population growth rates per 1000 people | -0.147 | 0.256 | 0.046 | 0.108 | -0.36 |
| Life expectancy at birth, years | 0.064 | 0.42 | 0.21 | 0.042 | 0.081 |
| Number of labor resources, thousand people | 0.304 | 0.09 | 0.32 | 0.175 | 0.258 |
| The share of persons under working age employed in the economy in the total number of employed | -0.103 | -0.06 | 0.108 | -0.113 | -0.089 |
| Unemployment rate | -0.212 | 0.467 | -0.151 | -0.08 | 0.14 |
| Real accounted wages of employees of organizations | -0.124 | 0.213 | -0.152 | 0.073 | 0.398 |

In order to validate the principal components model, the comparative hierarchical clustering procedure has been performed upon the initial dataset and the factor coordinates of observations (regions) in the obtained principal axes system. The complete linkage algorithm was applied with normal Euclidean distance as the metric (Fig. 7). It is readily observable that the cluster structure of the dataset has been preserved in the component modelling with minor differences, with clusters being somewhat more pronounced and alleviated.

The principal components axes can therefore be possibly interpreted as follows:

f1 – capital investments potential;

f2 – demographic load;

f3 – economic ability of population;

f4 – demographic capacity;

f5 – labor resources and labor reward (negative).

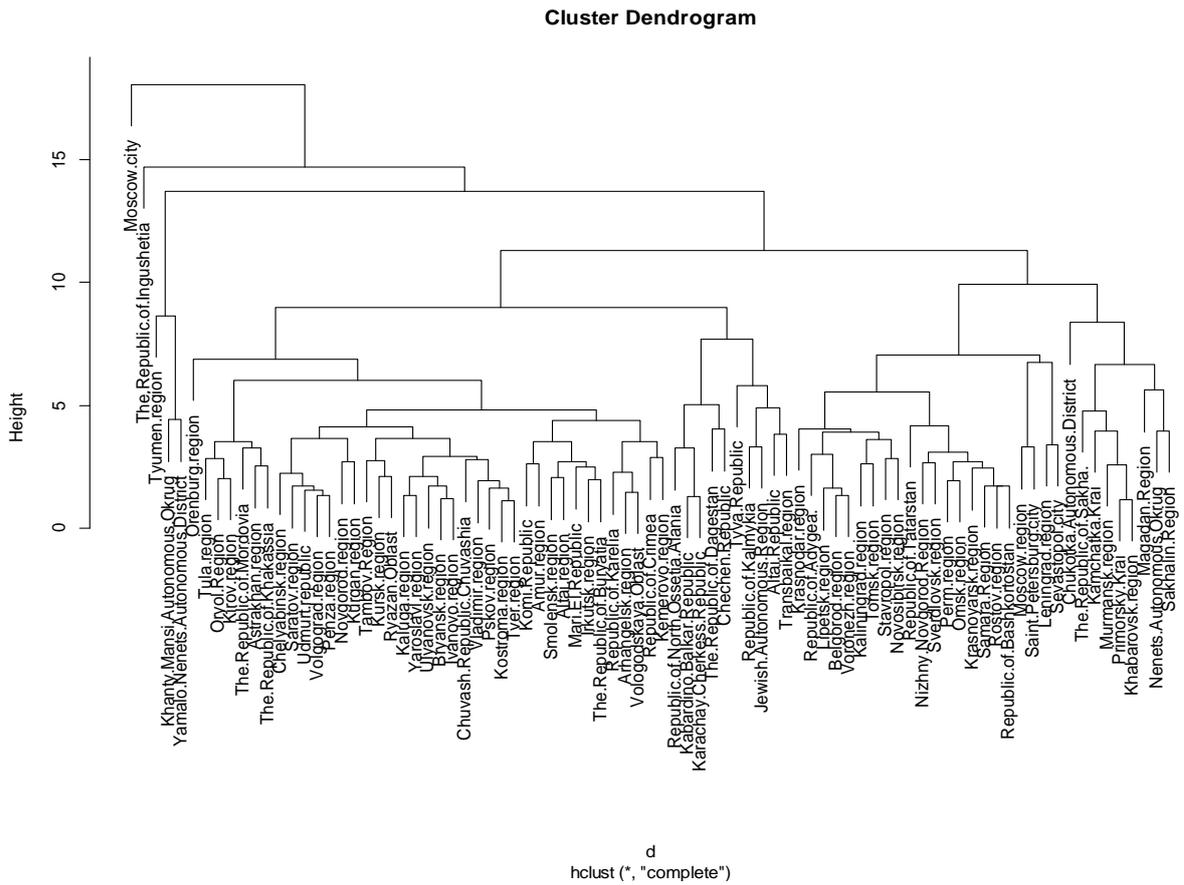

(a)

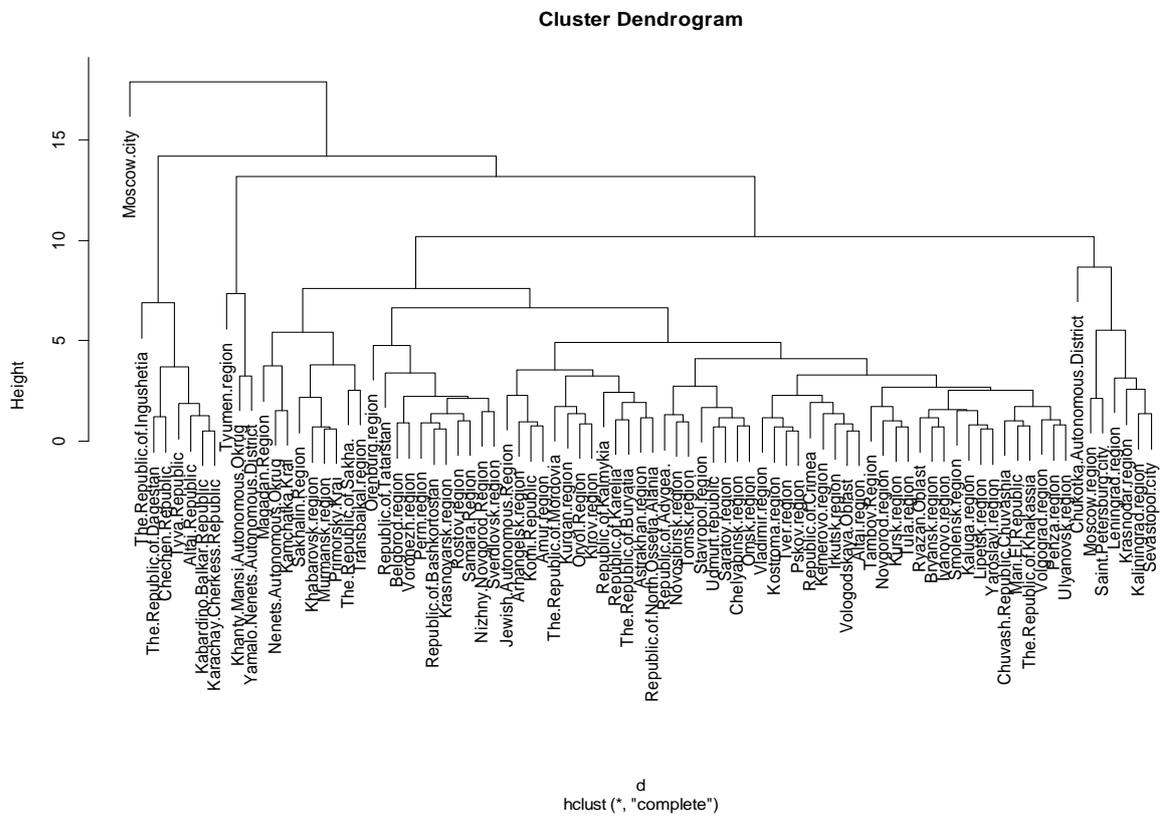

(b)

Figure 7: Hierarchical clusterization upon initial variables (a) and principal component scores (b)

The performed cluster analysis enables to identify four clusters among the regions of Russian Federation. Upon the regional clusterization, the structural and descriptive statistical analysis has been performed.

The first cluster is composed of the regions of Russia with the worst economic development and least provided population: The Republic of Dagestan, The Republic of Ingushetia, Kabardino-Balkar Republic, Karachay-Cherkess Republic, Chechen Republic, Altai Republic, and Tyva Republic. For these regions, the average per-capita cash income was 19864,86 rub., which is 42% below the country average. These regions are also characterized by the astoundingly small Gross Regional Product per capita and the volume of investments in fixed assets, which are 93,9% and 90,6% below the country average, respectively, and the highest proportions of population with cash incomes below subsistence level (24,44%). In terms of principal components analysis, these regions have the lowest capital investments potential (1,47 times below the country average), and the economic status of the population (1,2 times below the country average), as well as 2,96 and 2,4 times worse demographic situation and production capacity. However, these regions display certain availability of labor resources and their supplication, but are unable to effectively apply them, as the unemployment rate in these regions is 3 to 5 times greater than in other clusters (Table 4).

Table 4 - Low economy development regions of Russia

cluster characteristics, cluster average, 2018

| Indicator | Average | Standard deviation | Skewness | kurtosis | To country average, % |
|---|---|---|---|---|---|
| GRP PER CAPITA, RUBLES | 10822.51 | 8120.64 | 0.28 | -1.9 | -94% |
| THE VOLUME OF INVESTMENTS IN FIXED ASSETS, MILLION RUBLES | 45127.27 | 61940.45 | 1.37 | 0.21 | -91% |
| AVERAGE PER CAPITA CASH INCOME, RUBLES | 19864.86 | 3695.18 | 0.31 | -1.58 | -43% |
| THE SHARE OF THE POPULATION WITH CASH INCOMES BELOW THE SUBSISTENCE LEVEL, IN % | 24.44 | 6.43 | 0.11 | -1.3 | 63% |
| GINI COEFFICIENT, TIMES | 0.36 | 0.02 | 0.37 | -1.65 | -5% |
| CONSUMER PRICE INDEX , % | 103.16 | 0.67 | -0.72 | -1.05 | -1% |

The second cluster is composed of the regions of Russia with the highest economic development, including the regions of Moscow city, Tyumen region, Khanty-Mansi Autonomous District, and Yamalo-Nenets Autonomous District. These regions have ample economic potential by resource extraction (such as gold, coal, oil and gas), while Moscow city is the capital of Russian Federation. These regions are characterized by the GRP per capita 2.18 times and the investments in fixed assets 2.21 times greater than the country level; the average cash income per capita among these regions is 56915.75 rub. which is 64% above the country average and the highest among the four clusters. These regions also share the highest Gini coefficients in the country, the cluster average being 41%. The proportion of population below subsistence level in these regions is relatively small, amounting to 9.12% which is 39% below the country average. These regions are characterized with the highest production capacity and the highest labor reward (Table 5).

Table 5 – Highest economy development regions of Russia

cluster characteristics, cluster average, 2018

| Indicator | Average | Standard deviation | Skewness | kurtosis | To country average, % |
|---|---|---|---|---|---|
| GRP PER CAPITA, RUBLES | 562686.8 | 439499.9 | 0.35 | -1.88 | 218% |
| THE VOLUME OF INVESTMENTS IN FIXED ASSETS, MILLION RUBLES | 1547760 | 688442 | 0.02 | -2.4 | 221% |
| AVERAGE PER CAPITA CASH INCOME, RUBLES | 56915.75 | 21952.32 | -0.21 | -2.07 | 64% |
| THE SHARE OF THE POPULATION WITH CASH INCOMES BELOW THE SUBSISTENCE LEVEL, IN % | 9.12 | 4.08 | 0.53 | -1.85 | -39% |
| GINI COEFFICIENT, TIMES | 0.41 | 0.02 | 0.25 | -2.07 | 8% |
| CONSUMER PRICE INDEX , % | 102.9 | 1 | 0.47 | -1.88 | -1% |

The third cluster is constituted by the regions with high and mid-to-high economic development, and includes, notably, the Moscow Region, the St.Petersburg city and region, the Republic of Tatarstan, the Rostov region, the Samara region, the Republic of Sakha and some other. These regions have the Gini coefficient of 39%, the average cash income per capita being 35974.97 which is approximately at the country average level (4% higher). These regions, however, share smaller GRP per capita and smaller fixed capital investments (approximately below 50% of the country average). In terms of principal factors analysis, the economic ability among these regions is therefore better than average, but lower demographic capacity and labor rewards (the value of these principal factors is about 3 times lower than the country level).

It is worth noting that the regions in this cluster also share the highest skewness of fixed capital investments and the average cash income, as most people in these regions are characterized by below-cluster-average income levels (Table 6)

Table 6 – Mid-to-high economy development regions of Russia

cluster characteristics, cluster average, 2018

| Indicator | Average | Standard deviation | Skewness | kurtosis | To country average, % |
|---|---|---|---|---|---|
| GRP PER CAPITA, RUBLES | 91543.45 | 72769.43 | 0.93 | 0.45 | -48% |
| THE VOLUME OF INVESTMENTS IN FIXED ASSETS, MILLION RUBLES | 238388.7 | 203631.2 | 1.31 | 1.06 | -50% |
| AVERAGE PER CAPITA CASH INCOME, RUBLES | 35974.97 | 14943.3 | 1.7 | 2.02 | 4% |
| THE SHARE OF THE POPULATION WITH CASH INCOMES BELOW THE SUBSISTENCE LEVEL, IN % | 11.33 | 3.55 | 0.93 | 0.32 | -24% |
| GINI COEFFICIENT, TIMES | 0.39 | 0.02 | -0.22 | -0.9 | 3% |
| CONSUMER PRICE INDEX , % | 104.18 | 0.8 | -0.98 | 0.82 | 0% |

The fourth cluster includes the regions with mid and low-to-mid economic development, and includes the regions: Bryansk Region, Vladimir Region, Ivanovo Region, Kostroma Region, Kursk Region, Oryol Region, Ryazan District, Smolensk Region, Tver Region, Tula Region and others. The regions in this cluster are characterized by 25% below country average personal cash income (25915.36 rub.), and 15% population share with cash income below subsistence level which corresponds to the country average. The GRP per capita in these regions and the volume of investments in capital assets are 75% and 80% lower than country average, as well as lower demographic and labor resources capacity principal factors by 3.2 and 3.4 times. Within this cluster, the population income distribution is similar, as Gini coefficient equals to 36% among them and is slightly below the country average level (Table 7).

Table 7 – Low-to-mid economy development regions of Russia

cluster characteristics, cluster average, 2018

| Indicator | Average | Standard deviation | Skewness | kurtosis | To country average, % |
|---|---|---|---|---|---|
| GRP PER CAPITA, RUBLES | 43580.22 | 43137.65 | 2.51 | 7.03 | -75% |
| THE VOLUME OF INVESTMENTS IN FIXED ASSETS, MILLION RUBLES | 94954.83 | 74851.52 | 1.43 | 2.58 | -80% |
| AVERAGE PER CAPITA CASH INCOME, RUBLES | 25915.36 | 5553.53 | 1.29 | 1.61 | -25% |
| THE SHARE OF THE POPULATION WITH CASH INCOMES BELOW THE SUBSISTENCE LEVEL, IN % | 15.05 | 3.14 | 0.79 | 0.61 | 0% |
| GINI COEFFICIENT, TIMES | 0.36 | 0.02 | 0.53 | -0.4 | -5% |
| CONSUMER PRICE INDEX , % | 104.54 | 0.58 | 0.24 | -0.17 | 1% |

**4    Conclusions**

The results of analysis enable to state the following. The Russian Federation regions share distinctively non-uniform socio-economic development, and while the cluster structure is mixed due to high variation of the 'raw' parameters of territorial economic level, it is exposed better in performing the principal components analysis while maintaining the same cluster structure. The wealthiest regions of Russian Federation remain its capital and central ones, as well as the regions with core resources extraction industries (such as Moscow, St. Petersburg, Yamalo-Nenets, Tyumen, Chukotka Autonomous Districts). These regions share above-average population income and high capital investments, labor market competition is high within their territories and these regions are the general 'acceptors' of labor force.

The results of principal component modelling of Russian Federation regions by a set of 19 socio-economic parameters have enabled higher prominence of the cluster structure while allowing some space for certain variation of regional economic activity parameters. In conclusion, the proposed principal component model and the PCA procedure in general can be applied as a routine for cluster modelling of regional development.

Generalizing the results of the principal components modelling, makes it additionally possible to formulate the following conclusions: the other clusters aside from the afore-mentioned ones tend to be distinctively different from the highest-performing ones, lacking capital investment potential, the industrial growth points in them do not appear to have recovered from economic impacts that started between 2008 and 2015, including the sanctions that followed in 2014 and 2015, and while being competitive in terms of labor resources the economic potential in them shows signs of stagnation, and we are unlikely to see any economic improvement within them up until 2022-2025. The Covid-19 pandemic has further impacted the economic disparity existing within and among the regions of Russian Federation, and this poses further tasks for the government to alleviate its effects as well.

The income differentiation in the wealthiest regions is also likely to continue increasing; these regions are accumulating more wealth and the concentration of monetary and financial assets.

In addition upon the cluster analysis and the principal components analysis results it is possible to be observed that the Russian regions lack inter-territorial infrastructural integrity, as the neighbouring regions do not tend to achieve any synergic effect from being co-located and are diverse in terms of industrial areas.

The cluster analysis and regression modelling are heavily dependent upon exact similarities and differences by several parameters among the observations, which in certain occurrences may produce ambiguous or uncertain results when obtaining the regional cluster structure. This problem is even more evident where there exists the multicollinearity, while the initial parameters are numerous and their variation is distributed evenly among the studied regions.

**Acknowledgements**

This research was performed in the framework of the state task in the field of scientific activity of the Ministry of Science and Higher Education of the Russian Federation, project "Development of the methodology and a software platform for the construction of digital twins, intellectual analysis and forecast of complex economic systems", grant no. FSSW-2020-0008.

**References**

[1] Goal 10: Reduce inequality within and between countries. UN SDGs. - [Electronic resource]. - Access mode: https://www.un.org/sustainabledevelopment/ru/inequality/ (date accessed: 09/10/2020).

[2] On approval of the new edition of the state program "Provision of affordable and comfortable housing and utilities for citizens of the Russian Federation", available at: http://static.government.ru/media/files/v2A8bkUt5PQ.pdf

[3] The national project "Housing and Urban Environment", available at: http://static.government.ru/media/files/i3AT3wjDNyEgFywnDrcrnK7Az55RyRuk.pdf

[4] Federal project "Ensuring sustainable reduction of housing unsuitable for living", available at: https://storage.strategy24.ru/files/project/201904/feee652709d084b07996a8560ccaeaa2.pdf

[5] Comprehensive observation of living conditions of the population, available at: http://www.gks.ru/free_doc/new_site/KOUZ16/index.html

[6] Federal statistical observations on socio-demographic problems, available at: https://www.gks.ru/itog_inspect#

[7] Bezrukov A.V., Tenetova E.P. (2018), "Verification of the clustering of the innovative sector of the economy based on the inverse k-means procedure", Statistical studies of Russia's socio-economic development and prospects for sustainable growth: materials and reports, Plekhanov Russian University of Economics, Moscow, Russia, 2018, pp. 26-32

[8] State Corporation - Fund for Assistance to the Reform of Housing and Communal Services. Analytics, available at: https://www.reformagkh.ru/analytics#overhaul-section

[9] Davletshina L.A. (2017), "Analysis of the development of the regions of the Russian Federation according to the main indicators of the socio-demographic situation", Materials of the 3rd International Correspondence Scientific and Practical Conference "Statistical Analysis of the Socio-Economic Development of the Subjects of the Russian Federation", Bryansk, Russia, 2017, pp. 78-82

[10] Dedyukhina ES, Petrenko M.A. Foreign experience in the overhaul of apartment buildings using innovative mechanisms, available at: https://cyberleninka.ru/article/n/zarubezhnyy-opyt-kapitalnogo-remonta-mnogokvartirnyh-domov- s-ispolzovaniem-innovatsionnyh-mehanizmov

[11] Dolgikh EA, Pershina T.A. (2019), "Studying Cross-Country Differences in Innovation Potential", working paper, Trends in the Development of Science and Education, May 2018

[12] Housing and communal services Control. Teaching aids. Housing Education, available at: http://gkhkontrol.ru/2015/09/26286

[13] Makhova O.A., Karmanov M.V., Arakelyan S.M. (2018), "Statistics as a digitalization tool", Statistical studies of the socio-economic development of Russia and prospects for sustainable growth: materials and reports, Plekhanov Russian University of Economics, Moscow, Russia, 2018, pp. 174-178

[14] Pershina T.A., Davletshina L.A. (2018), "Statistical analysis of a generalized integral indicator of the socio-economic situation of the constituent entities of the Russian Federation", working paper, Vestnik of the University, State University of Management, May 2018

[15] Filippov V.S., Vashchekina I.V., Komarova I.P. et. al. (2015), "Results of operational monitoring of socio-economic development of Russia and the regions of Russian Federation", Results of operational monitoring of socio-economic development of Russia and the regions of Russian Federation, Plekhanov Russian University of Economics, Situational center for social and economic development of regions, Moscow, Russia, 2015, vol. 8.

[16] Ermolaev S.A., Komarova I.P., Sigarev A.V. et al. "Analytical notes. Results of operational monitoring of socio-economic development of Russia and the regions of Russian Federation", Plekhanov Russian University of Economics, Situational center for social and economic development of regions, Moscow, Russia, 2017, vol. 26.